\documentclass[paper]{ieice}
\usepackage[dvipdfmx]{graphicx,xcolor}
\usepackage[fleqn]{amsmath}
\usepackage{newtxtext}
\usepackage[varg]{newtxmath}
\usepackage{enumerate}

\newcommand{\cv}[1]{{\langle{#1}\rangle}}
\newcommand{\lengthx}[1]{\vert #1 \vert}

\def\plane{{\mathcal R}^2}

\def\net{{\mathcal N}}
\def\link{{\mathcal L}}

\def\path{{\mathcal P}}

\def\discon{\not\leftrightarrow}
\def\outer{{\mathcal O}}
\def\bq{\begin{equation}}
\def\eq{\end{equation}}
\def\bqn{\begin{eqnarray}}
\def\eqn{\end{eqnarray}}
\newtheorem{theorem}{\bf{Theorem}}

\newtheorem{col}{\bf{Corollary}}

\field{B}
\title{Theoretical Analysis for Determining Geographical Route of Cable Network with Various Disaster-Endurance Levels}
\authorlist{%
 \authorentry{Hiroshi Saito}{f}{UTokyo}\MembershipNumber{8402516}
}
\affiliate[UTokyo]{The author is with the Mathematics and Informatics Center, The University of Tokyo.}

\received{0}{0}{0}
\revised{0}{0}{0}

\begin{document}
\maketitle

\begin{summary}
This paper theoretically analyzes cable network disconnection due to randomly occurring natural disasters, where the disaster-endurance (DE) levels of the network are determined by a network entity such as the type of shielding method used for a duct containing cables. The network operator can determine which parts have a high DE level. When a part of a network can be protected, the placement of that part can be specified to decrease the probability of disconnecting two given nodes.

The maximum lower bound of the probability of connecting two given nodes is explicitly derived.  Conditions decreasing (not decreasing) the probability of connecting two given nodes with a partially protected network are provided.

\end{summary}

\section{Introduction}
The world has been impacted by a large number of severe natural disasters, such as earthquakes, tsunamis, and hurricanes, which take thousands of human lives and destroy network infrastructures \cite{AsBeSo11}. For example, a severe earthquake in March 2011 off the northeast coast of Japan and its associated tsunami killed many people and destroyed facilities, including network facilities, in cities and towns \cite{ntt_commag}. Earthquakes that cause similar damage occur every few years worldwide, such as the Shichuan earthquake in China in 2008 \cite{china}, \cite{survey0}. The damage caused by an earthquake is huge and has a global impact. 

Network infrastructure is becoming increasingly important, and the destruction of a network seriously impacts society. Therefore, network operators do their best to minimize damage to their networks from natural disasters. Typical disaster countermeasures are based on protection, prompt restoration, and securing critical communications such as 911. Examples of protection are building disaster-resistant facilities and preparing backup systems, and those focused on prompt restoration are introducing mobile equipment such as a transportable terrestrial station for a satellite communication system \cite{ntt}. However, service disruption will inevitably occur due to devastating natural disasters. It is therefore necessary to improve the robustness of networks against such disasters in general and earthquakes in particular through new and current methods \cite{commag_article}.

The recently proposed ^^ ^^ disaster-free network'' \cite{disaster-free} is a concept completely different from others based on protection and restoration. It aims at avoiding disasters as much as possible and is implemented through disaster avoidance control \cite{INFOCOMdisaster_avoidance}, \cite{ToN_disaster_avoidance} and physical network design \cite{ToN_saito}, \cite{JLT_saito}, \cite{Commag_Tran}, \cite{JLT_Tran}, \cite{infocomSaito}. The former dynamically changes the geographical shape of a network, and the latter determines the geographical/geometrical shape of a physical network.

This paper proposes a theoretical method for determining the geographical/geometrical shape of a cable network along the concept of the disaster-free network. The contributions of this paper are as follows.

This paper investigates a network with multiple disaster-endurance (DE) levels that are determined by a network entity such as the type of shielding method used for a duct containing cables. The network operator can then determine which parts have a high DE level.  When a part of a network can be protected, the placement of that part is specified so as to decrease the probability of two given nodes being disconnected.   In addition, the maximum lower bound of the probability of connecting two given nodes is explicitly derived.  Conditions decreasing (not decreasing) the probability of connecting two given nodes with a partially protected network are provided.

The organization of this paper is as follows. Section 2 presents related work. Section 3 provides the model and notations used in this paper . In Section 4, some of the results in previous works are presented to use them in this paper. Section 5 provide the main results. Numerical examples are given in Section 6, and the paper is concluded in Section 7.

\section{Related work}
Many papers have been published on network-protection and service-restoration methods \cite{survey0}. This section focuses on the geographical/geometrical design methods of a physical network.

Geographical design methods use geographical information, such as terrain and geological features or the frequency of earthquakes of each geographical area, to determine the geographical routes of a network. 
Mathematical optimization with some constraints is often used for these methods.
For example, an earthquake hazard map of Japan was used to optimally reconfigure routes of existing cables \cite{Commag_Tran} and derive optimal geographical routes of newly installed ducts/cables \cite{JLT_Tran}. 
To determine the geographical route of an undersea cable based on the estimated likelihood of an earthquake, Zhao et al. \cite{zukerman_taiwan} solved a graph optimization problem to obtain an optimal solution that balances cost and survivability. 
In \cite{protection}, a multi-objective optimization solution that takes into account the cost of laying optical fibers and repair cost with various cable-protection methods was investigated. This solution has been applied to determine the geographical route and protection methods to be used.
Recently, when a geographical route information of a power grid is given, a geographical design method of a network using the power grid is proposed \cite{second_type}.   In that paper, geographical areas are divided into sub-areas of multiple disaster-vulnerable levels.

Geometrical assumptions have also been introduced to make the optimization model simple or to derive an explicit solution. For example, a disaster area is modeled as a disk, half plane, or finite convex area, while the geographical network shape is assumed to be, for example, a rectangle. 
A disaster area was modeled as a strip or half plane and the probability of disconnecting two nodes was explicitly derived in \cite{ToN_saito} and that for a probabilistic failure was derived in \cite{infocomSaito}. Saito \cite{JLT_saito} modeled a disaster area as a finite convex area and determined the optimal geographical/geometrical shape of the route of ducts/cables. In \cite{underseaCableFailure} the optimal route of an undersea cable was investigated by assuming a disk-shaped disaster area.  Assuming a rectangular route makes it possible to determine the length of an edge by minimizing the cost. Cao et al. \cite{light} extended that study to other route shapes. 

Some studies have evaluated network survivability under certain geometrical assumptions. Directly designing the geographical shape of a network may not be possible, but network survivability can be evaluated for various geographical shapes of a network. Gardner et al. \cite{cqr} also considered a disk-shaped disaster model to analyze the connectivity between the source and destination. Neumayer et al. published two papers on network survivability in a disaster \cite{failureToN}, \cite{failureINFOCOM}. Their network model is a set of line segments where the end points are locations of network center buildings. The disaster model is a line segment or circle \cite{failureToN}. They proposed an algorithm to identify worst-case disasters. 

There have also been studies on minimum-cut-max-flow. To the best of my knowledge, Bienstock \cite{OR} initiated the study of this problem. Algorithms for computing the minimum number of disaster areas disconnecting the source and sink node were investigated when all the edges intersecting in a disaster areas were removed. 
Sen et al. \cite{sen} proposed region-based connectivity as a metric for fault tolerance. Based on the assumption that the region is a disk-shaped disaster area, polynomial time algorithms for calculating region-based connectivity were developed. Neumayer et al. \cite{discMinCut} discussed the geographical min-cut, defined as the minimum number of disk-shaped disaster areas necessary to disconnect a pair of nodes, and the geographical max-flow, defined as the maximum number of paths that are not disconnected by a single disaster area. The important finding in that study is that geographical min-cut is not equal to geographical max-flow. Agarwal et al. studied algorithms for finding a disaster location that has the highest expected impact on a network, where the impact is defined with various metrics such as the number of failed components \cite{wdmFailure}. Zhang et al. \cite{icct} evaluated the risk of each region by searching for the worst line-cut. Trajanovski et al. \cite{3pointFit} also studied this and proposed a polynomial time algorithm for finding a critical region and a region-disjoint path. This algorithm is based on the finding that three points can determine the location of an elliptical or polygon disaster area.

\section{Model and notations}
\subsection{Model}
Let $\net$ be an optical fiber cable network between $s$ and $t$ within a bounded and convex $\Omega$, which is an area of interest. Disasters causing damage in part of $\Omega$ are taken into account. We are interested in $\Pr(s\leftrightarrow t)$, the probability of connecting between $s$ and $t$ during a disaster, or $\Pr(s\discon t)=1-\Pr(s\leftrightarrow t)$, the probability of disconnecting $s$ and $t$ due to a disaster. There may be multiple geographical cable routes between $s$ and $t$ in $\net$.

The disaster area $D\subset \plane$ is modeled as a randomly placed area around $\net (\subset \Omega\subset \plane)$. In the remainder of this paper, it is assumed that a disaster area $D$ is geographically much larger than $\net$. For example, the $D$ of a large earthquake is at least hundreds of km$^2$. Some may reach tens of thousands of km$^2$. A large hurricane can create a disaster area larger than a hundred km$^2$. Therefore, this assumption is useful, for example, for evaluating a disaster affecting a regional network or for designing a robust physical route of such a network against disasters. 

Because $D$ is very large, we can assume that its boundary is macroscopically a line (the validity of this model was verified through the numerical results using field data provided by \cite{ToN_saito}). For a directional line $G$, $D$ is assumed to be $R_G$, which means the right-half plane of $G$. That is, $D=R_G$. When we analyze the geographical shape of a network to reduce the possibility of encountering the disaster, $R_G$ completely including $\Omega$ is meaningless because the network of any shape is contained in the disaster area. The $R_G$ not intersecting $\Omega$ is also meaningless because the network of any shape is not contained in the disaster area. For our objective, we should focus on cases $G\cap \Omega \neq \emptyset$ and assume $G\cap \Omega \neq \emptyset$ in the remainder of this paper.

\begin{figure}[tbh] 
\begin{center} 
\includegraphics[width=8cm]{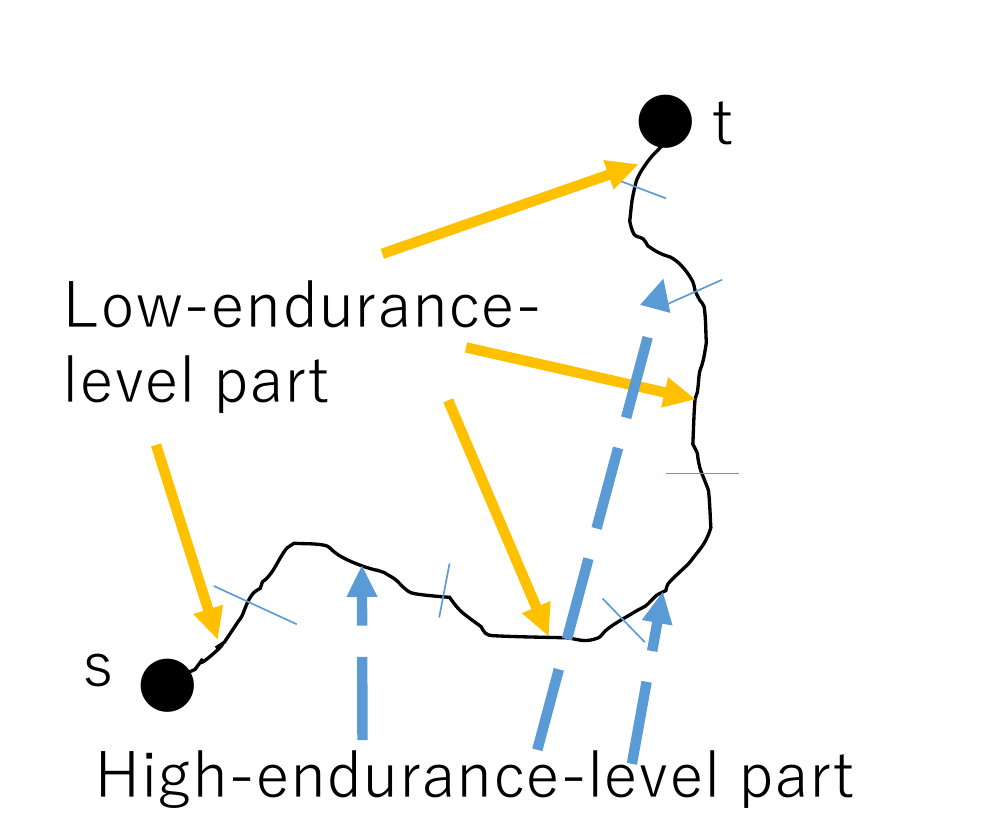}
\caption{Illustration of disaster-endurance levels.}
\label{first_typeDE}
\end{center} 
\end{figure}

$D$ has a disaster level $L$ denoting the intensity of the disaster. It can relate to the intensity of an earthquake, the wind speed of a tornado, etc. A part of $\net$ is characterized by a DE level. A part of $\net$ with a DE level $L$ means that it is destroyed and disconnected for a level-$L$ disaster or higher but is not destroyed for a disaster with a lower level than $L$. The level is determined by a network entity such as the type of the shielding method used for a duct containing cables \cite{protection} (Fig. \ref{first_typeDE}). The network operator can thus determine which parts have a high DE level. 

\subsection{Notations}
Let $\cv{X}$, $\lengthx{X}$, and $\partial X$ be the convex hull, perimeter length, and boundary of $X\subset\plane$, respectively. The notation $\cv{X_1,\cdots,X_k}$ means the convex hull of $X_1\cup\cdots\cup X_k$. $X^c$ is the complement set of $X$. For two points $x,y\subset \plane$, $l(x,y)$ is the line segment between them. The notation $m(X)$ is the measure of the sets of lines satisfying $X$. (The measure is proportional to the probability. By normalizing the measure, it becomes the probability, which is called ^^ ^^ geometric probability'' \cite{Santalo},\cite{saitoHP}.)

\begin{table}[h]
\caption{List of notations.}
\begin{center}
\begin{tabular}{ll}
\hline
$\cv{X}$&convex hull of $X\subset\plane$\cr
$\lengthx{X}$&perimeter length of $X\subset\plane$\cr
$\partial X$&boundary of $X\subset\plane$\cr
$X^c$&complement set of $X$\cr
$l(x,y)$& line segment between $x,y\subset \plane$\cr
$m(X)$&measure of sets of lines satisfying $X$ \cr
$D$& disaster area\cr 
$\net$&physical network\cr
$\Omega$& area of interest\cr
$\Pr(s\leftrightarrow t)$&probability of maintaining connectivity between $s$ and $t$\cr
$\Pr(s\discon t)$&probability of disconnecting $s$ and $t$\cr
		&$\Pr(s\discon t)=1-\Pr(s\leftrightarrow t)$\cr
$G$&directional line\cr
$R_G$&right-half plane of $G$, $D=R_G$ in this paper\cr
$\net_\Phi(L)$&set of parts of $\net$ that are destroyed by a level-$L$ disaster\cr
$\outer_i$&$i$-th outer route\cr
$I_i(\net)$&set of inner parts of $\net$ in $\outer_i$\cr
$L,\{L_i\}_i$& disaster level\cr
\hline
\end{tabular}
\end{center}
\end{table}

\section{Preliminary}
\begin{figure}[htb] 
\begin{center} 
\includegraphics[width=6cm,clip]{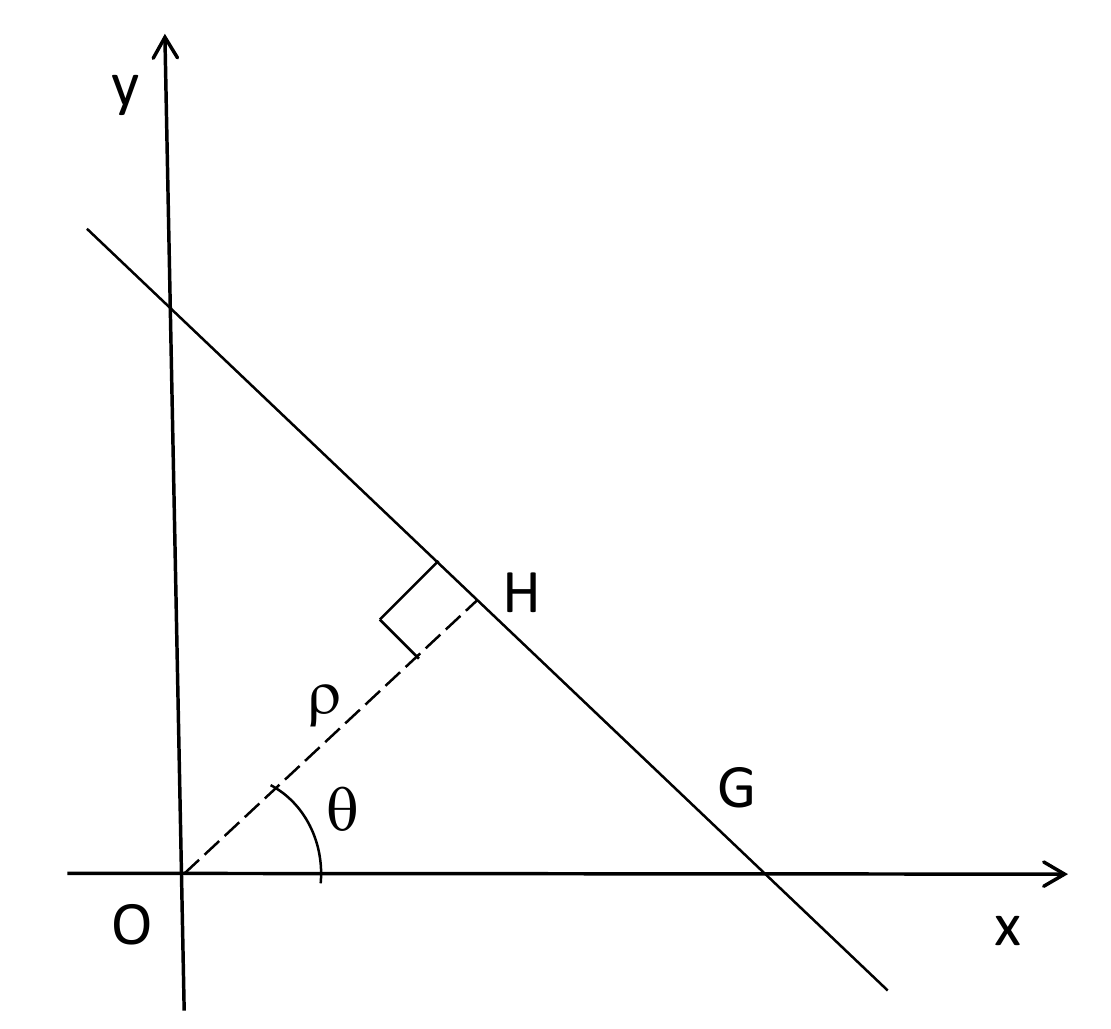} 
\caption{Parameterization of line.} 
\label{lineset} 
\end{center} 
\end{figure}

The concept of integral geometry and geometric probability \cite{Santalo} is introduced here as a preliminary for evaluating a disaster occurring at a random location. We can define the measure of a set of lines. Consider a line $G$ determined by the angle $\theta$ and by its distance $\rho$ from the origin $O$ ($0\leq \rho$).
The angle $\theta$ is made with the direction perpendicular to $G$ and the positive part of the $x$-axis ($-\pi \leq \theta \leq \pi$)  (Fig. \ref{lineset}). That is, $G$ is specified by the coordinates $(\rho,\theta)$. The motion-invariant measure of the set of lines $G(\rho,\theta)$ satisfying $X$ is defined by the simple integral form $m(X)=\int_X d\rho\,d\theta$ \cite{Santalo}. Throughout this paper, any boundary of a set in $\plane$ is smooth and differentiable except for the finite number of points.

Equation (3.12) in \cite{Santalo} gives
\bq\label{3.12}
m(G\cap C\neq\emptyset)=\lengthx{C}
\eq
for a convex set $C$.

In general, for sets $X\subseteq Y$, $\Pr(X|Y)$ can be defined as follows \cite{Santalo}.
\bq\label{def}
\Pr(X|Y)=m(X)/m(Y)
\eq

The following are the results of Theorem 1 and Eq. (3) in \cite{ToN_saito}: Let $C$ be a set in $\plane$ and assume that $C\subset\Omega$, where $\Omega$ is bounded and convex. Then 
\bq\label{pre}
\Pr(R_{G} \cap C= \emptyset\vert G\cap \Omega\neq \emptyset)=\frac{\lengthx{\Omega}-\lengthx{\cv{C}}}{2\lengthx{\Omega}}.
\eq

\section{Analysis}
This section analyzes and yields $\Pr(s\leftrightarrow t)$ for a network of multiple DE levels. 
Theorem \ref{Type1th2} presents that the weakest arrangement defined below maximizes $\Pr(s \discon t)=1-\Pr(s\leftrightarrow t)$.
For a single-route network, $\Pr(s\leftrightarrow t)$ is given as a function of a set of weak parts of $\net$ through Theorem \ref{th1}. 
For a network of multiple routes, the closed form $\Pr(s\leftrightarrow t)$ under the weakest arrangement is derived in Theorem \ref{th_m1}.
Theorem \ref{protect} provides how the placement of protected parts increases or does not increase $\Pr(s\leftrightarrow t)$.

For a level-$L$ disaster, define $\net_\Phi(L)$ as the set of parts of $\net$ that are destroyed and disconnected (we may remove $L$ for simplicity).

\begin{figure}[tbh] 
\begin{center} 
\includegraphics[width=8cm]{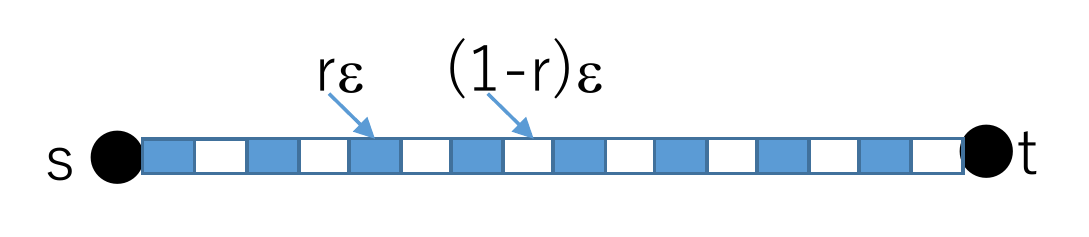}
\caption{Weakest arrangement.}
\label{weakest}
\end{center} 
\end{figure}

The weakest arrangement for a level-$L$ disaster is defined as follows. (The arrangement means a placement method of protected parts and non-protected parts.) Assume we can protect $100r\%$ ($0<r<1$) of the network from a level-$L$ disaster. Divide each link in $\net$ by a pair of parts with lengths $r\epsilon, (1-r)\epsilon$, where $\epsilon\rightarrow 0$ (Fig. \ref{weakest}). That is, each link in $\net$ alternatively consists of a part disconnected and a part not disconnected by a level-$L$ disaster. This arrangement on $\net$ is defined as the weakest arrangement. (For simplicity, ^^ ^^ for a level-$L$ disaster" may be removed.)

\begin{theorem}\label{Type1th2}
For a level-$L$ disaster area $D$, $\Pr(s\discon t)$ is maximized under the weakest arrangement.
\end{theorem}
{\bf Proof:} When a location of $D$ is given, a set $S_\link$ of links intersecting $D$ is determined, where a link $\link\in S_\link$ satisfies $\link\cap D \neq \emptyset$.
$\link\cap D$ always includes a part disconnected under the weakest arrangement. 
Thus, $\link\in S_\link$ is always disconnected under the weakest arrangement.
That is, $S_\link$ is a set of links disconnected due to $D$ under the weakest arrangement.

Let $S'_\link$ be a set of links disconnected due to $D$ under any arrangement.
A link $\link$ disconnected due to $D$ satisfies $\link\cap D \neq \emptyset$.
Thus, $S'_\link\subseteq S_\link$.
This means that $\Pr(s\discon t)$ under the weakest arrangement is larger than or equal to that under any arrangement.
$\square$

\subsection{Single-route network}
This subsection discusses the case in which $\net$ consists of a single route between $s$ and $t$.

\begin{theorem}\label{th1}
For a level-$L$ disaster, $\Pr(s\leftrightarrow t)$ is given as follows.
\bq
\Pr(s\leftrightarrow t)=\frac{\lengthx{\Omega}-\lengthx{\cv{\net_\Phi(L)}}}{2\lengthx{\Omega}}.
\eq
$\net_\Phi(L)$ minimizing $\lengthx{\cv{\net_\Phi(L)}}$ maximizes $\Pr(s\leftrightarrow t)$.
\end{theorem}
{\bf Proof:} Note that $\Pr(s\leftrightarrow t)=\Pr(\net_\Phi\cap D=\emptyset|G\cap \Omega \neq \emptyset)$. Under the condition that $D=R_G$, $\Pr(\net_\Phi\cap R_G=\emptyset|G\cap \Omega \neq \emptyset)=\frac{\lengthx{\Omega}-\lengthx{\cv{\net_\Phi}}}{2\lengthx{\Omega}}$ due to Eq. (\ref{pre}).

It is clear that $\Pr(s\leftrightarrow t)$ is maximized when $\lengthx{\cv{\net_\Phi}}$ is minimized.
$\square$

\subsection{Multiple routes}
This subsection discusses the case in which $\net$ has more than one route between $s$ and $t$.

\begin{figure}[tbh] 
\begin{center} 
\includegraphics[width=8cm]{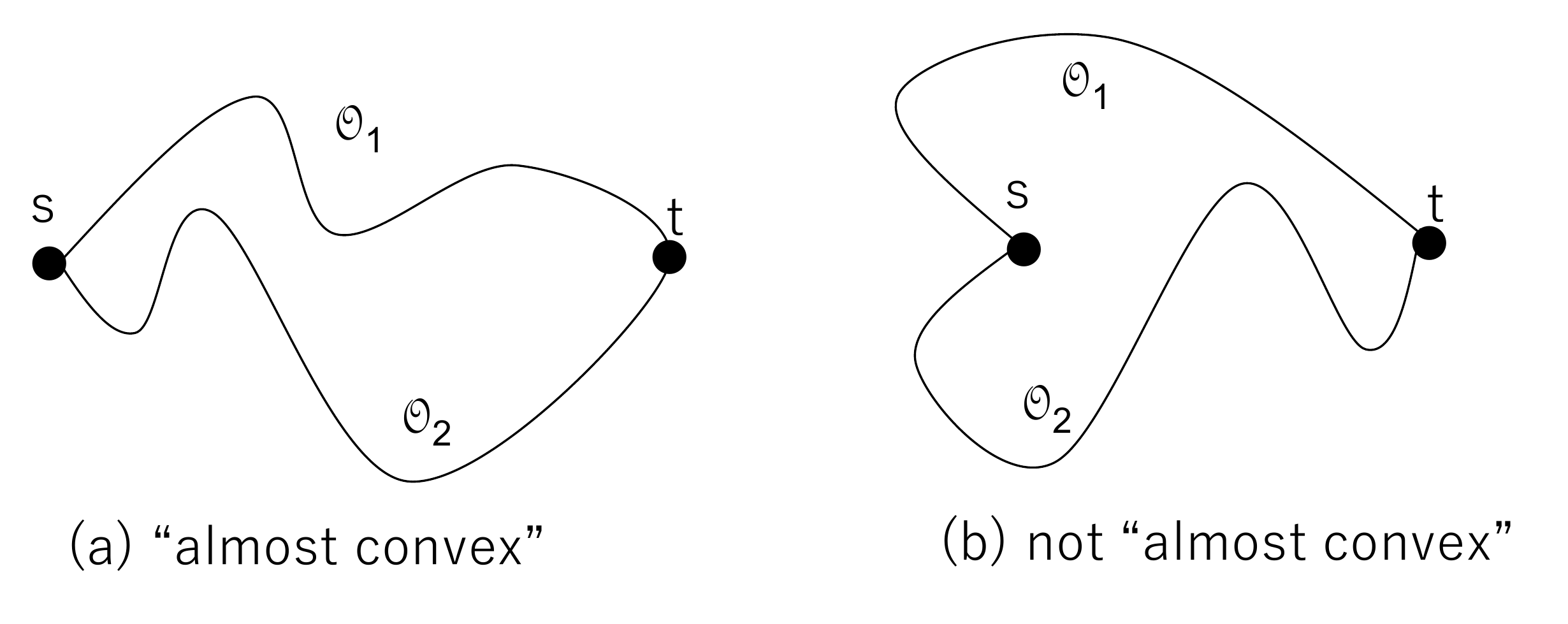}
\caption{Illustration of almost-convex network.}
\label{local-concave}
\end{center} 
\end{figure}

\begin{figure}[tbh] 
\begin{center} 
\includegraphics[width=8cm]{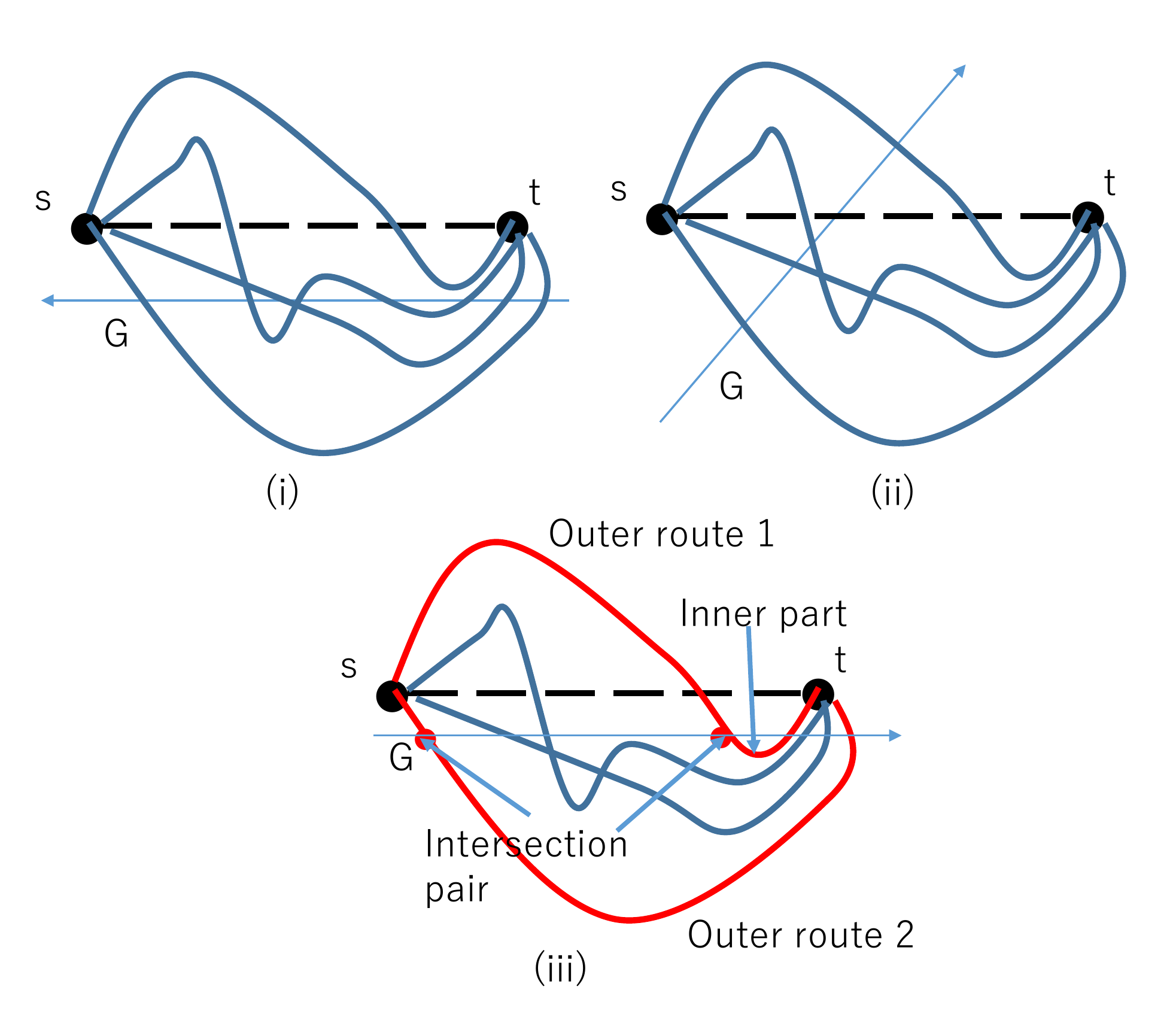}
\caption{Events (i)--(iii) for Theorem \ref{th_m1}.}
\label{multiple_routes}
\end{center} 
\end{figure}

Assume there are two routes called outer routes among the routes in $\net$. Each outer route does not intersect or overlap with any other routes (except at $s,t$), and the area enclosed by the two outer routes contains all the routes in $\net$ (Fig. \ref{multiple_routes}). When a part of an outer route is between the other outer route and $l(s,t)$, this part is called an inner part. For each half plane defined by the line passing through $s$ and $t$, there may be inner parts. For example, in Fig. \ref{multiple_routes}-(iii), the upper half plane does not have an inner part while the lower half plane has one inner part. Let $I_i(\net)$ be the set of inner parts of $\net$ in the $i$-th outer route $\outer_i$. When $s,t\subset\partial\cv{\outer_1\cup\outer_2}$, we call the network ^^ ^^ almost convex'' (Fig. \ref{local-concave}). Note that when a network $\outer_1\cup\outer_2$ is convex, it is almost convex. If $\outer_1\cup\outer_2$ is not almost convex, $G$ may intersect both $\outer_1$ and $\outer_2$ without intersecting $l(s,t)$ or the inner parts. In the remainder of this subsection, assume that there exist two outer routes in $\net$.

\begin{theorem}\label{th_m1}
Under the weakest arrangement, $\Pr(s\leftrightarrow t)$ is given as follows when $\net$ is almost convex and there exist multiple routes between $s$ and $t$.
\bqn\label{th_m1eq}
&&\Pr(s\leftrightarrow t)\cr
&=&\frac{\lengthx{\Omega}+\lengthx{l(s,t)}-\sum_{i=1}^2\lengthx{\cv{I_i(\net),l(s,t)}}}{2\lengthx{\Omega}}.
\eqn

\end{theorem}
{\bf Proof:} We should first note that, under the weakest arrangement, any part of a route includes a part of $\net_\Phi$. Hence, a part of $\net$ in $R_G$ is always disconnected under the weakest arrangement.

Let us prove Eq. (\ref{th_m1eq}) under the assumption of the weakest arrangement.

\begin{figure}[tbh] 
\begin{center} 
\includegraphics[width=8cm]{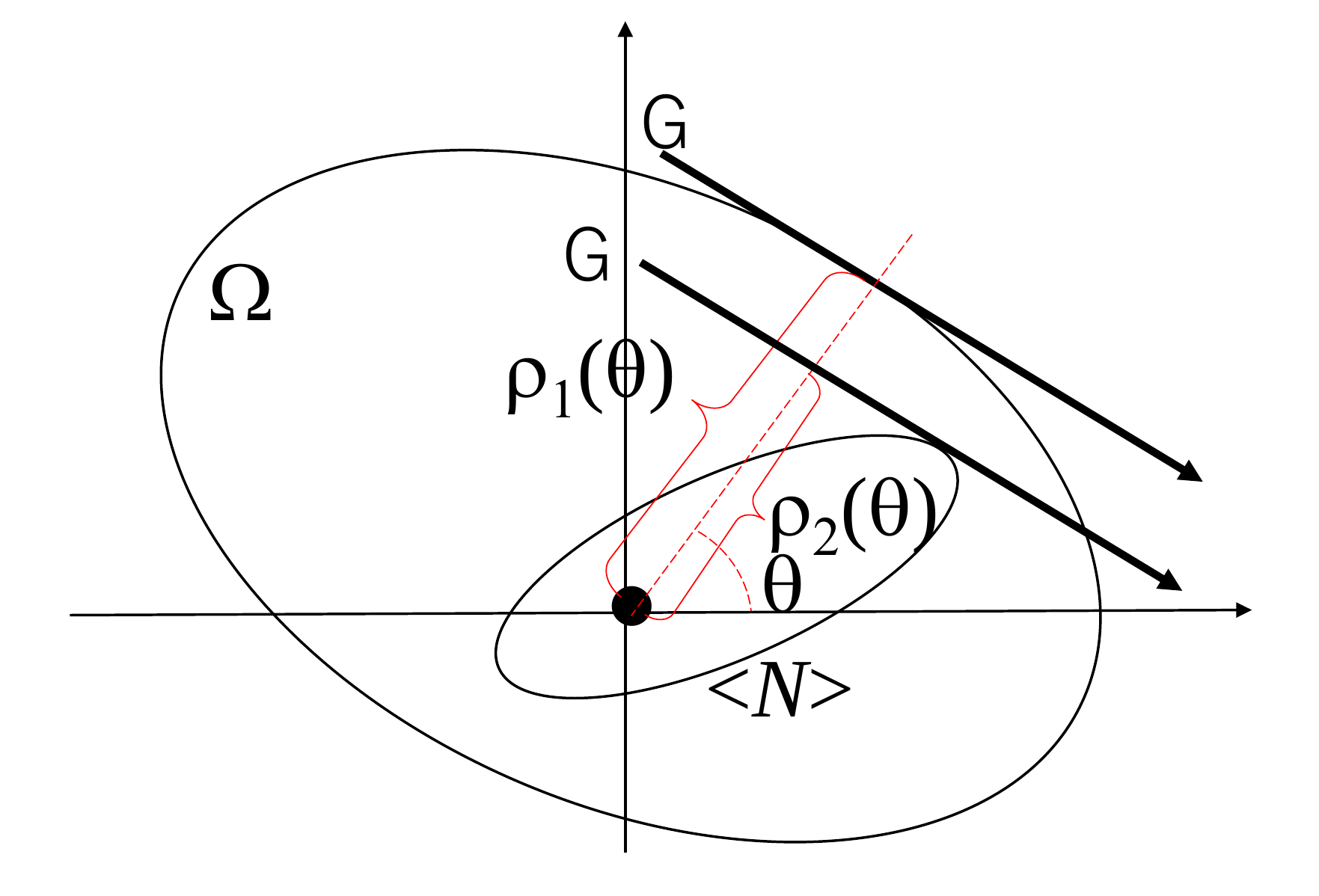}
\caption{Illustration of $m(\net\subset R_G,\net\cap G=\emptyset, G\cap \Omega\neq \emptyset)$.}
\label{measure}
\end{center} 
\end{figure}

When $G$ satisfying $G\cap \Omega\neq \emptyset$ satisfies $\{\net\subset R_G,\net\cap G=\emptyset\}$, none of the routes between $s$ and $t$ work. For a fixed $\theta$, the range of $\rho$ that satisfies $\net\subset R_G,\net\cap G=\emptyset, G\cap \Omega\neq \emptyset$ is $\rho_2\leq \rho\leq \rho_1$ if the direction of $G$ is set to cover $\net\subset R_G$ (Fig. \ref{measure}). When $\rho=\rho_1$, $G$ becomes a tangent line (formally, a supporting line) of $\Omega$. When $\rho=\rho_2$, $G$ becomes a tangent line (formally, a supporting line) of $\cv{\net}$. Thus, 
\bqn
&&m(\net\subset R_G,\net\cap G=\emptyset, G\cap \Omega\neq \emptyset)\cr
&=&\int_{-\pi}^{\pi}\int_{\rho_2}^{\rho_1}(1/2)d\rho\ d\theta\cr
&=&\int_{-\pi}^{\pi}(\rho_1-\rho_2)/2\,d\theta,
\eqn
where $1/2$ is needed because there are two possibilities that $\net$ is contained in the right- or left-half plane of $G$. When $\rho(\theta)$ is a supporting function for $X$, the following relationship is known between $\rho(\theta)$ and its perimeter length $\lengthx{X}$: $\int_{-\pi}^\pi \rho(\theta) d\theta=\lengthx{X}$ \cite{Santalo}. Thus,
\bq
m(\net\subset R_G,\net\cap G=\emptyset, G\cap \Omega\neq \emptyset)=(\lengthx{\Omega}-\lengthx{\cv{\net}})/2.
\eq
Because $m(G\cap \Omega\neq \emptyset)=\lengthx{\Omega}$ due to Eq. (\ref{3.12}), 
\bqn
&&\Pr(\net\subset R_G,\net\cap G=\emptyset|G\cap \Omega\neq \emptyset)\cr
&=&\frac{m(\net\subset R_G,\net\cap G=\emptyset, G\cap \Omega\neq \emptyset)}{m(G\cap \Omega\neq \emptyset)}\cr
&=&(\lengthx{\Omega}-\lengthx{\cv{\net}})/(2\lengthx{\Omega}).
\eqn
The first equality is due to Eq. (\ref{def}).

When $G$ satisfying $G\cap \Omega\neq \emptyset$ satisfies $\net\cap G\neq\emptyset$, disconnection of all routes between $s$ and $t$ is equivalent to the occurrence of one of the following exclusive events for all the routes: (i) both a part around $s$ and a part around $t$ are in $R_G$, (ii) either a part around $s$ or around $t$ is in $R_G$ and the other is not in $R_G$, and (iii) neither the part around $s$ nor that around $t$ is in $R_G$ but a part in the middle of each route is in $R_G$ (Fig. \ref{multiple_routes}).

First, consider event (i). Note that the measure of the set of $G$ satisfying event (i) is the half of the measure of the set of $G$ that $G\cap \net\neq\emptyset, G\cap l(s,t)=\emptyset$. This is because both a part around $s$ and a part around $t$ are in the right- or left-half plane of $G$ when $G\cap \net\neq\emptyset, G\cap l(s,t)=\emptyset$. Thus, 
\bqn\label{eq(i)}
&&m(G \,\rm{satisfying}\, (i))\cr
&=&m(G\cap \net\neq\emptyset, G\cap l(s,t)=\emptyset)/2\cr
&=&(m(G\cap \net\neq\emptyset)-m(G\cap \net\neq\emptyset, G\cap l(s,t)\neq\emptyset))/2\cr
&=&(m(G\cap \net\neq\emptyset)-m(G\cap l(s,t)\neq\emptyset))/2\cr
&=&(\lengthx{\cv{\net}}-\lengthx{l(s,t)})/2.
\eqn
The third equality uses the fact that $G\cap \net\neq\emptyset$ if $G\cap l(s,t)\neq\emptyset$. The fourth equality uses Eq. (\ref{3.12}).

Second, note that event (ii) is equivalent to the event that $G\cap l(s,t)\neq\emptyset$. $G\cap\Omega\neq\emptyset$ is satisfied if $G\cap l(s,t)\neq\emptyset$. Then, using Eq. (\ref{3.12}),
\bq\label{eq(ii)}
m(G \,\rm{satisfying}\, (ii))=\lengthx{l(s,t)}.
\eq

Now consider event (iii). Focus on a half plane determined by a line passing through $s$ and $t$. 
If there is no inner part, $G$ cannot intersect all the routes in a half plane because there exists an outer route outside that half plane.

Assume there is an inner part(s) of an outer route in a half plane in the remainder of this proof. To intersect the outer route containing the inner part under event (iii), $G$ needs to intersect (the convex hull of) the inner part without intersecting $l(s,t)$. Conversely, $G$ intersecting the convex hull of the inner part and not intersecting $l(s,t)$ intersects all the routes in the half plane. This is because (a) the area enclosed by the inner part (parts), the other outer route, and $l(s,t)$ contain all the routes in the half plane, and (b) $G$ makes an intersection pair in the half plane where the pair has an intersection, A, of $G$ and the convex hull of the inner part and the other intersection, B, of $G$ and the other outer route (Fig. \ref{multiple_routes}-(iii)). Note that all the routes in the half plane intersect $G$ between A and B. Thus, event (iii) is equivalent to $G$ intersecting the convex hull of inner parts and not intersecting $l(s,t)$. Because there are two possibilities, namely, that a half plane contains both $s$ and $t$ or neither, 
\bqn
&&m(G \,\rm{satisfying}\, (iii))\cr
&=&\sum_{i=1}^2 m(G\cap\cv{I_i(\net)}\neq\emptyset,G\cap l(s,t)=\emptyset)/2.
\eqn
Note that the event $G\cap (I_i(\net)\cup l(s,t))\neq\emptyset$ is equivalent to the following three exclusive events: $\{G\cap I_i(\net)\neq\emptyset,G\cap l(s,t)\neq\emptyset\}$, $\{G\cap I_i(\net)\neq\emptyset, G\cap l(s,t)=\emptyset\}$, and $\{G\cap I_i(\net)=\emptyset,G\cap l(s,t)\neq\emptyset\}$. In addition, $G\cap I_i(\net)=(\neq)\emptyset$ is equivalent to $G\cap\cv{I_i(\net)}=(\neq)\emptyset$, and $G\cap (I_i(\net)\cup l(s,t))=(\neq)\emptyset$ is equivalent to $G\cap \cv{I_i(\net), l(s,t)}=(\neq)\emptyset$. Thus, 
\bqn
&&m(G\cap\cv{I_i(\net)}\neq\emptyset,G\cap l(s,t)=\emptyset)\cr
&=&m(G\cap\cv{I_i(\net),l(s,t)}\neq\emptyset)\cr
&&\qquad-m(G\cap I_i(\net)\neq\emptyset,G\cap l(s,t)\neq\emptyset)\cr
&&\qquad-m(G\cap I_i(\net)=\emptyset,G\cap l(s,t)\neq\emptyset)\cr
&=&m(G\cap\cv{I_i(\net),l(s,t)}\neq\emptyset)-m(G\cap l(s,t)\neq\emptyset)\cr
&=&\lengthx{\cv{I_i(\net),l(s,t)}}-\lengthx{l(s,t)}.
\eqn
The last equality uses Eq. (\ref{3.12}).

As a result,
\bq\label{eq(iii)}
m(G \,\rm{satisfying}\, (iii))=\sum_{i=1}^2(\lengthx{\cv{I_i(\net),l(s,t)}}-\lengthx{l(s,t)})/2.
\eq

Consequently, 
\bqn
&&\Pr(s\leftrightarrow t)\cr
&=&1-(\Pr(\net\subset R_G,\net\cap G=\emptyset|G\cap \Omega\neq \emptyset)\cr
&&\qquad+\sum_{j=(i),(ii),(iii)}m(G \,\rm{satisfying}\, j)/m(G\cap\Omega\neq\emptyset))\cr
&=&\frac{\lengthx{\Omega}+\lengthx{l(s,t)}-\sum_{i=1}^2\lengthx{\cv{I_i(\net),l(s,t)}}}{2\lengthx{\Omega}}.
\eqn
$\square$

Theorem \ref{th_m1} suggests that $\Pr(s\leftrightarrow t)$ is independent of the geographical shape of the non-inner parts of the routes if $\net$ is almost convex and the weakest arrangement is used. This means that the effectiveness of the changes in the routes is limited, at least under the weakest arrangement. In addition, due to Theorem \ref{Type1th2}, Eq. (\ref{th_m1eq}) gives the lower bounds of $\Pr(s\leftrightarrow t)$ (upper bounds of $\Pr(s\discon t)$).

\begin{figure}[tbh] 
\begin{center} 
\includegraphics[width=8cm]{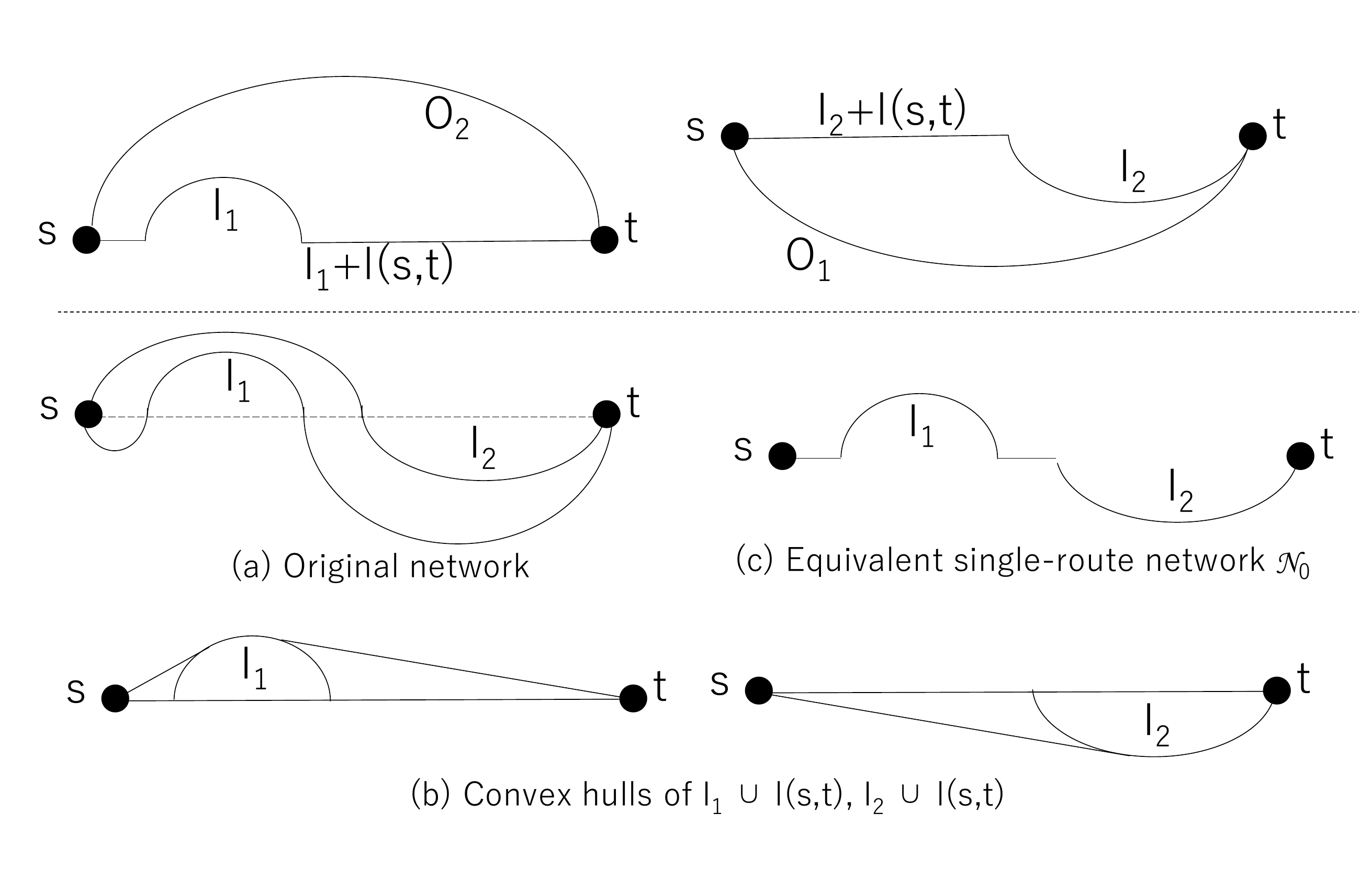}
\caption{Equivalent single-route network.}
\label{combined-multi}
\end{center} 
\end{figure}

Let $\outer_i=I_i+ l(s,t)$ denote that $\outer_i$ consists of its inner part $I_i$ and a line segment (segments) connecting $I_i$, $s$, and $t$, where $i=1,2$.  See the upper two figures in Fig. \ref{combined-multi}.

\begin{col}\label{col0}
Under the weakest arrangement, if $\outer_1=I_1+ l(s,t)$ or $\outer_2=I_2+ l(s,t)$, $\Pr(s\leftrightarrow t)$ is given by Eq. (\ref{th_m1eq}).
\end{col}
{\bf Proof:} When $\outer_1=I_1+ l(s,t)$ and $s,t$ are in the left-half plane of $G$, the disconnection of $\outer_1$ means the disconnection of the inner part of $\outer_1$. When the inner part is disconnected, $\outer_2$ is also disconnected. Thus, similarly to the proof of Theorem \ref{th_m1}, $\Pr(s\leftrightarrow t)$ is given by Eq. (\ref{th_m1eq}). $\square$ 

We should note that Corollary \ref{col0} does not need the assumption that $\net$ is ^^ ^^ almost convex.''
This corollary is used in the numerical example for the network in Fig. \ref{realNet}.

\begin{col}\label{col1}
Under the weakest arrangement, for $\net$ that is almost convex and has multiple routes between $s$ and $t$, there exists a single-route network $\net_0$ of which $\Pr(s\leftrightarrow t)$ is equal to that of $\net$, where $\net_0$ consists of $I_1(\net),I_2(\net)$ and the line segments connecting them and $s$ and $t$.
\end{col}
{\bf Proof:} Note that $\sum_{i=1}^2\lengthx{\cv{I_i(\net),l(s,t)}}=\lengthx{\cv{\net_0}}+\lengthx{l(s,t)}$ (Fig. \ref{combined-multi}). Thus, according to Eq. (\ref{th_m1eq}), $\Pr(s\leftrightarrow t)$ of $\net$ is 
\bqn
&&\Pr(s\leftrightarrow t)=\frac{\lengthx{\Omega}-\lengthx{\cv{\net_0}}}{2\lengthx{\Omega}}.
\eqn
Because of Eq. (\ref{pre}), this probability is $\Pr(s\leftrightarrow t)$ of $\net_0$ under the weakest arrangement.
$\square$

The following corollary provides a stronger corollary than the one mentioned above, because the former means the latter but the reverse is not always true.

\begin{col}\label{col2}
Under the weakest arrangement, for $\net$ that is almost convex and has multiple routes between $s$ and $t$, $\net_0$ is equivalent to $\net$ in the following sense.  Disconnecting  (connecting) $s$ and $t$ through $\net_0$ means disconnecting  (connecting) $s$ and $t$ through $\net$, and conversely, disconnecting  (connecting) $s$ and $t$ through $\net$ means disconnecting  (connecting) $s$ and $t$ through $\net_0$.
\end{col}
{\bf Proof:} First, assume disconnecting $s$ and $t$ through $\net_0$ and prove disconnecting $s$ and $t$ through $\net$. Disconnecting $s$ and $t$ through $\net_0$ means one of the following exclusive events occurs: (i) $\{s\subset R_G\}\cup\{t \subset R_G\}$, (ii) $\{s,t\not\subset R_G\}\cap(\{I_1\cap R_G\neq \emptyset\}\cup\{I_2\cap R_G\neq \emptyset\})$. For (i), $s$ and $t$ are disconnected even when $\net$ is used. For (ii), if $I_1\cap R_G\neq \emptyset$ ($I_2\cap R_G\neq \emptyset$), then $\outer_2\cap R_G\neq \emptyset$ ($\outer_1\cap R_G\neq \emptyset$) because $I_1$ and $\outer_2$ ($I_2$ and $\outer_1$) do not intersect. Thus, disconnecting $s$ and $t$ through $\net_0$ results in disconnecting $s$ and $t$ through $\net$.

Next, assume disconnecting $s$ and $t$ through $\net$ and prove disconnecting $s$ and $t$ through $\net_0$.
Disconnecting $s$ and $t$ through $\net$ means (i) $\{s\subset R_G\}\cup\{t \subset R_G\}$ and (ii) $G$ intersects $I_i$ and its outer part of $\outer_j$ ($(i,j)=(1,2),(2,1)$) and $s,t\not\subset R_G$.
((ii) corresponds to the proof of Theorem \ref{th_m1} (iii).)
For (i), it is clear that $\net_0$ is disconnected.
For (ii), because $I_i$ is disconnected, $\net_0$ is disconnected.
$\square$

Now, a partial protect arrangement of $\net$ is defined as follows. For a given disaster level, a continuous part $\Gamma\subset \outer_1\cup\outer_2$ is not destroyed and the remaining parts are destroyed.

\begin{figure}[tbh] 
\begin{center} 
\includegraphics[width=8cm]{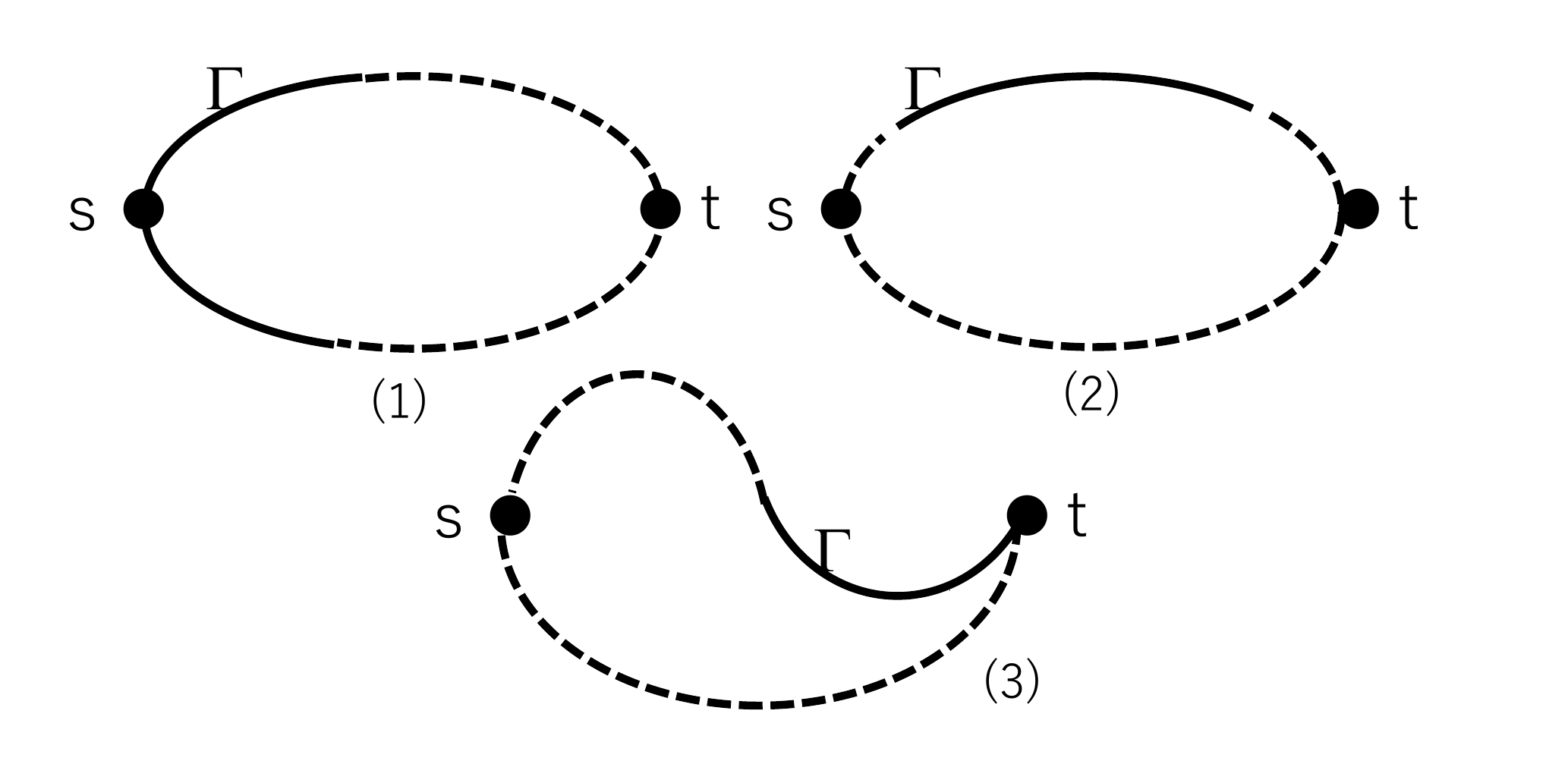}
\caption{Illustration of Theorem \ref{protect}.}
\label{partial_protect}
\end{center} 
\end{figure}

\begin{theorem}\label{protect}
For $\Pr(s\leftrightarrow t)$ under the weakest arrangement, $\Pr(s\leftrightarrow t)$ under a partial protect arrangement is (1) improved if $\Gamma$ contains $s$ or $t$, (2) not improved if $\outer_1,\outer_2$ have no inner parts, $\net$ is almost convex, and $\Gamma$ does not contain $s$ or $t$, or (3) improved if $\net$ is almost convex and there exists an inner part contained in $\Gamma$ (Fig. \ref{partial_protect}).
\end{theorem}

{\bf Proof:} For (1), assume that $\Gamma$ contains $s$ or $t$. Without loss of generality, we can assume that $\Gamma$ contains $s$. Then, there exists a line $G_0\cap l(s,t)\neq\emptyset$ where $G_0$ intersects only in $\Gamma$. For $s\subset R_{G_0}$, $\outer_1$ and $\outer_2$ maintain connectivity under a partial protect arrangement while they are disconnected under the weakest arrangement. It is clear that disconnection between $s$ and $t$ under a partial protect arrangement occurs under the weakest arrangement. Therefore, $\Pr(s\leftrightarrow t)$ is improved under a partial protect arrangement.

For (2), assume that both $\outer_1$ and $\outer_2$ are disconnected under the weakest arrangement. Their disconnections mean that (i) both $s$ and $t$ are in $R_G$, or (ii) either $s$ or $t$ is in $R_G$ and the other is not (because $\outer_1,\outer_2$ have no inner parts and $\net$ is almost convex, disconnecting $s,t\not\subset R_G$ cannot occur). 
For (i), the network is disconnected under both arrangements because $\Gamma$ does not contain $s$ or $t$. Now we concentrate on (ii) and note that it is equivalent to the event that $G\cap l(s,t)\neq\emptyset$. If $G\cap l(s,t)\neq\emptyset$, $s$ or $t$ included in $R_G$ is not protected even under a partial protect arrangement because of the assumption that $\Gamma$ does not contain $s$ or $t$. Therefore, the network is disconnected and $\Pr(s\leftrightarrow t)$ is not improved.

For (3), there exists a line $G_0$ that intersects $I_1\subset\Gamma$ and does not intersect $l(s,t)$. Then, $\outer_1$ containing $I_1$ is not disconnected under a partial protect arrangement with probability of 1/2 because both $s,t$ are contained in the left-half plane of $G_0$ with probability 1/2 (under the weakest arrangement, $I_1$ is disconnected and $s$ and $t$ are disconnected even when $s,t$ are contained in the left-half plane of $G_0$). It is clear that disconnection between $s$ and $t$ under a partial protect arrangement occurs under the weakest arrangement. Therefore, $\Pr(s\leftrightarrow t)$ is improved under a partial protect arrangement.
$\square$

This theorem demonstrates the effectiveness of making the network robust around the source or destination.
In addition, making inner parts robust improves network survivability. This is because the disconnection of inner parts, which can occur even under $G\cap l(s,t)=\emptyset$, results in the disconnection of the network. This fact suggests that the parts not making a network almost convex should be protected to improve the network survivability. This is because disconnection of these parts can occur under $G\cap l(s,t)=\emptyset$ and results in the disconnection of the network. This suggestion is verified in a numerical example (Fig. \ref{non-convex-protect}).

\section{Numerical examples}
In this section, numerical examples are presented. Here, $\Omega$ is a disk with the radius of $r_{\Omega}$ and a center at the origin. In the simulation, $10^6$ disasters were randomly located to obtain a sample.

\subsection{Simple example}
\begin{figure}[tbh] 
\begin{center} 
\includegraphics[width=8cm]{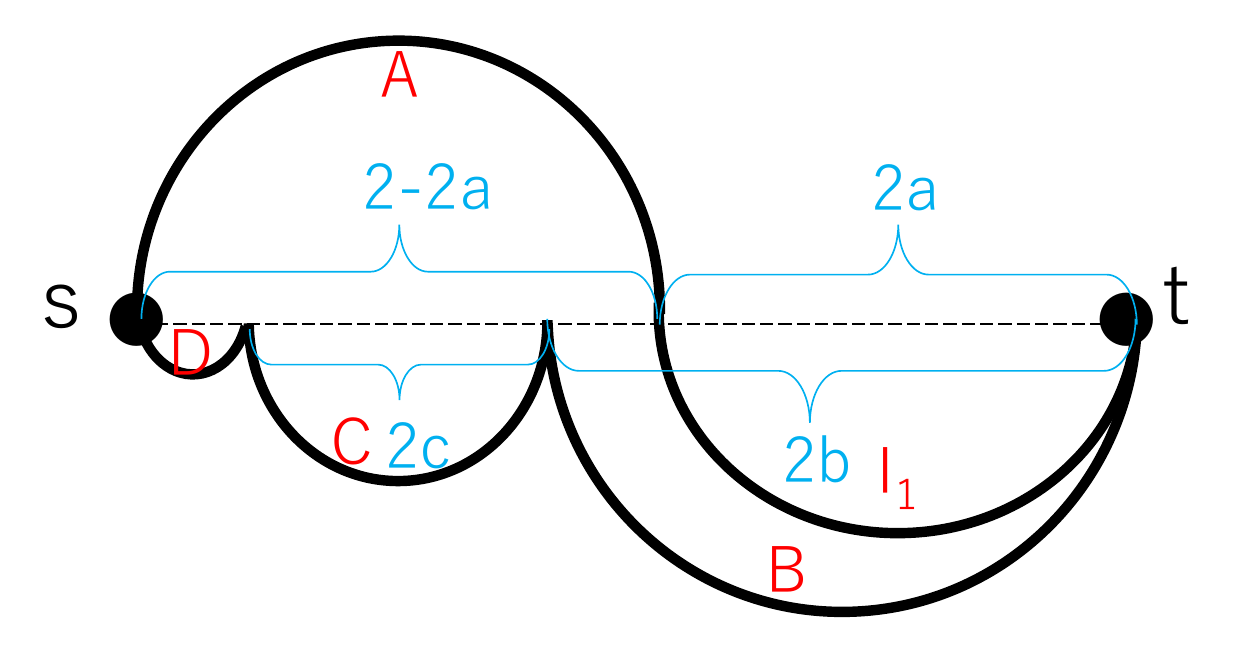}
\caption{Example 1.}
\label{example1}
\end{center} 
\end{figure}

The example network we used is shown in Fig. \ref{example1}. There are two routes between $s$ and $t$. The first route consists of half circles $I_1$ and A with radii of $a,1-a$. The second route consists of half circles B, C, and D with radii of $b,c,1-b-c$. The distance ($\lengthx{l(s,t)}/2$) between $s$ and $t$ is 2, and $r_{\Omega}=2$. The middle point of $l(s,t)$ is at the origin. This network was used to evaluate the results.

\begin{figure}[tbh] 
\begin{center} 
\includegraphics[width=8cm]{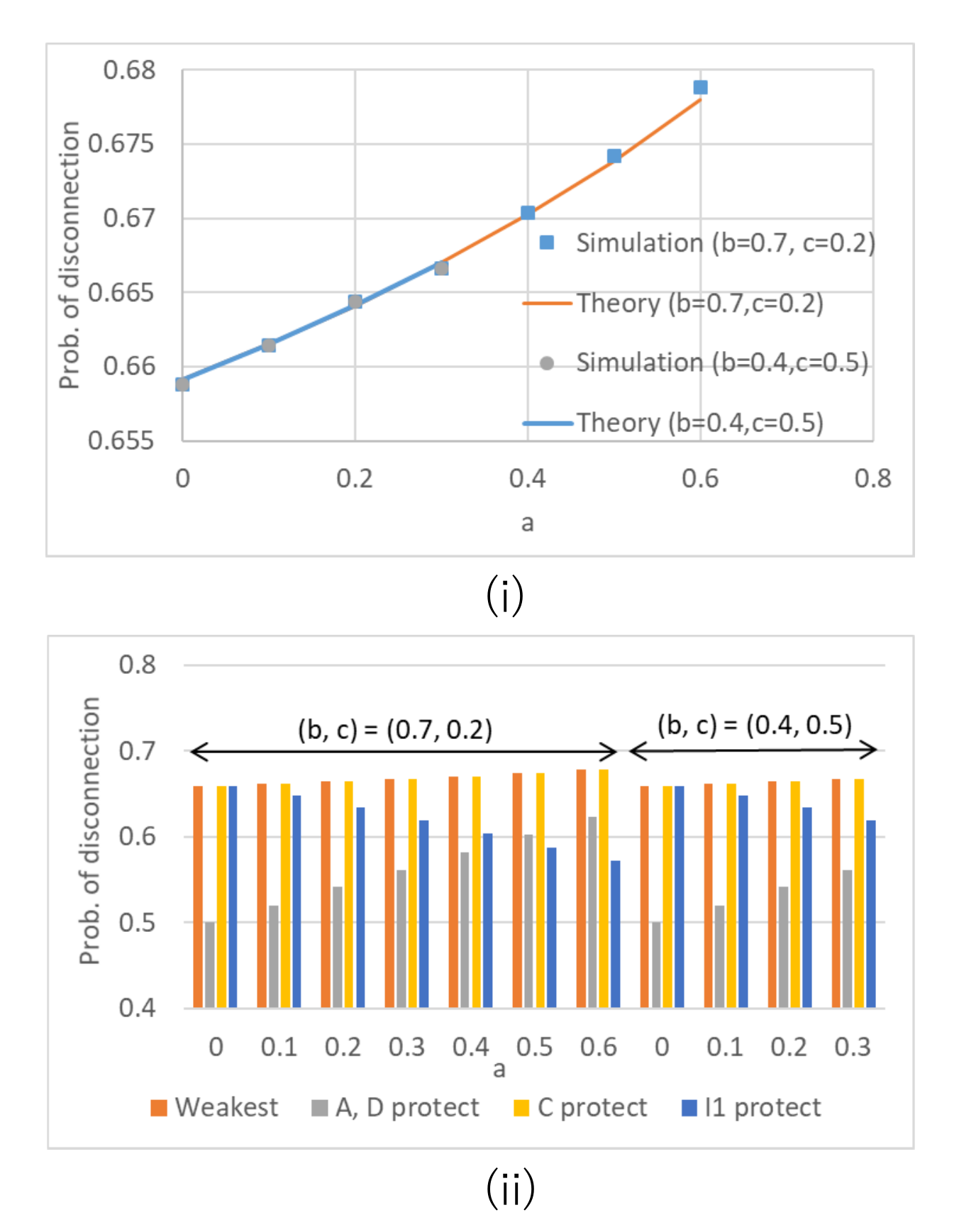}
\caption{Results of example 1.}
\label{example1a_result}
\end{center} 
\end{figure}

Figure \ref{example1a_result} provides the results of Theorems \ref{th_m1} and \ref{protect}.

In Fig. \ref{example1a_result}-(i), $\Pr(s \discon t)=1-\Pr(s\leftrightarrow t)$ is plotted from the simulation and Eq. (\ref{th_m1eq}), where $\lengthx{\cv{I_1,l(s,t)}}=\sqrt{(\lengthx{l(s,t)}/2-a)^2-a^2}+a(\pi-\arccos(a/(\lengthx{l(s,t)}/2-a))+\lengthx{l(s,t)}/2$ and $\lengthx{\cv{I_2,l(s,t)}}=\lengthx{l(s,t)}$ in Fig. \ref{example1}. 
Figure \ref{example1a_result}-(i) demonstrates the followings. (1) $\Pr(s \discon t)$ grows as $a$ increases, that is, the inner part becomes larger. (2) $\Pr(s \discon t)$ is independent of $b$ because Eq. (\ref{th_m1eq}) is independent of $b$. Although there is no graph, it is also independent of $c$. This is because Eq. (\ref{th_m1eq}) is independent of $c$.

Figure \ref{example1a_result}-(ii) also plots $\Pr(s \discon t)$ by using the simulation to confirm Theorem \ref{protect}. According to Theorem \ref{protect}, compared with the weakest arrangement, (1) $\Pr(s \discon t)$ decreases when A and D are protected because $s$ is protected, (2) $\Pr(s \discon t)$ does not decrease when $a=0$ and C is protected, and (3) $\Pr(s \discon t)$ decreases when $I_1$ is protected. These three properties are confirmed in this figure.

\begin{figure}[tbh] 
\begin{center} 
\includegraphics[width=8.5cm]{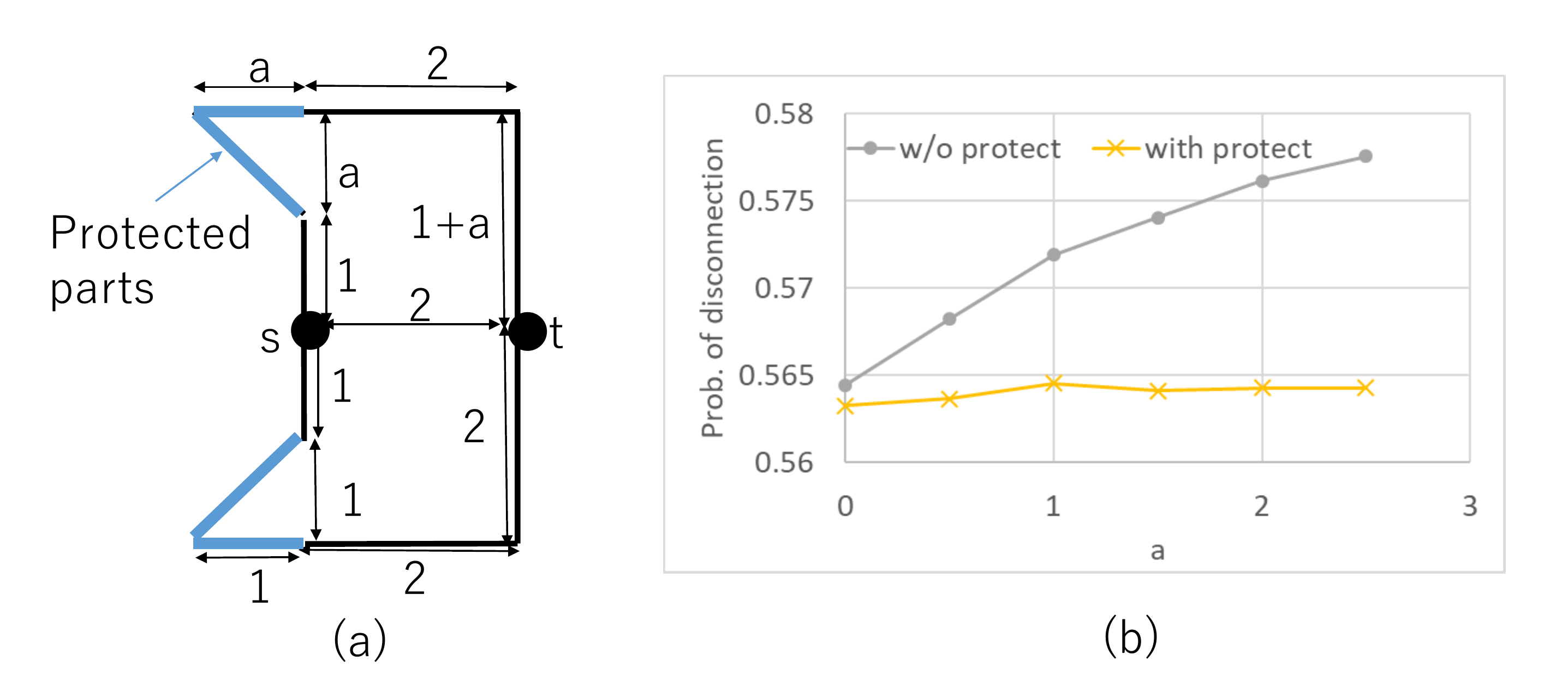}
\caption{Network model for extending Theorem \ref{protect} and its simulation results.}
\label{non-convex-protect}
\end{center} 
\end{figure}

Theorem \ref{protect} suggests that the parts not making the network almost convex should be protected to improve network survivability. To confirm this suggestion, the following example is provided. The network used in this example is shown in Fig. \ref{non-convex-protect}-(a). The origin is at the middle point between $s$ and $t$, and $r_\Omega=5$. The protected part is in blue.

The simulation results are presented in Fig. \ref{non-convex-protect}-(b). As expected, the protection decreased $\Pr(s\discon t)$, although the protected parts did not contain $s$, $t$, or inner parts.

\subsection{Realistic network model}
\begin{figure}[tbh]
\begin{center} 
\includegraphics[width=8cm]{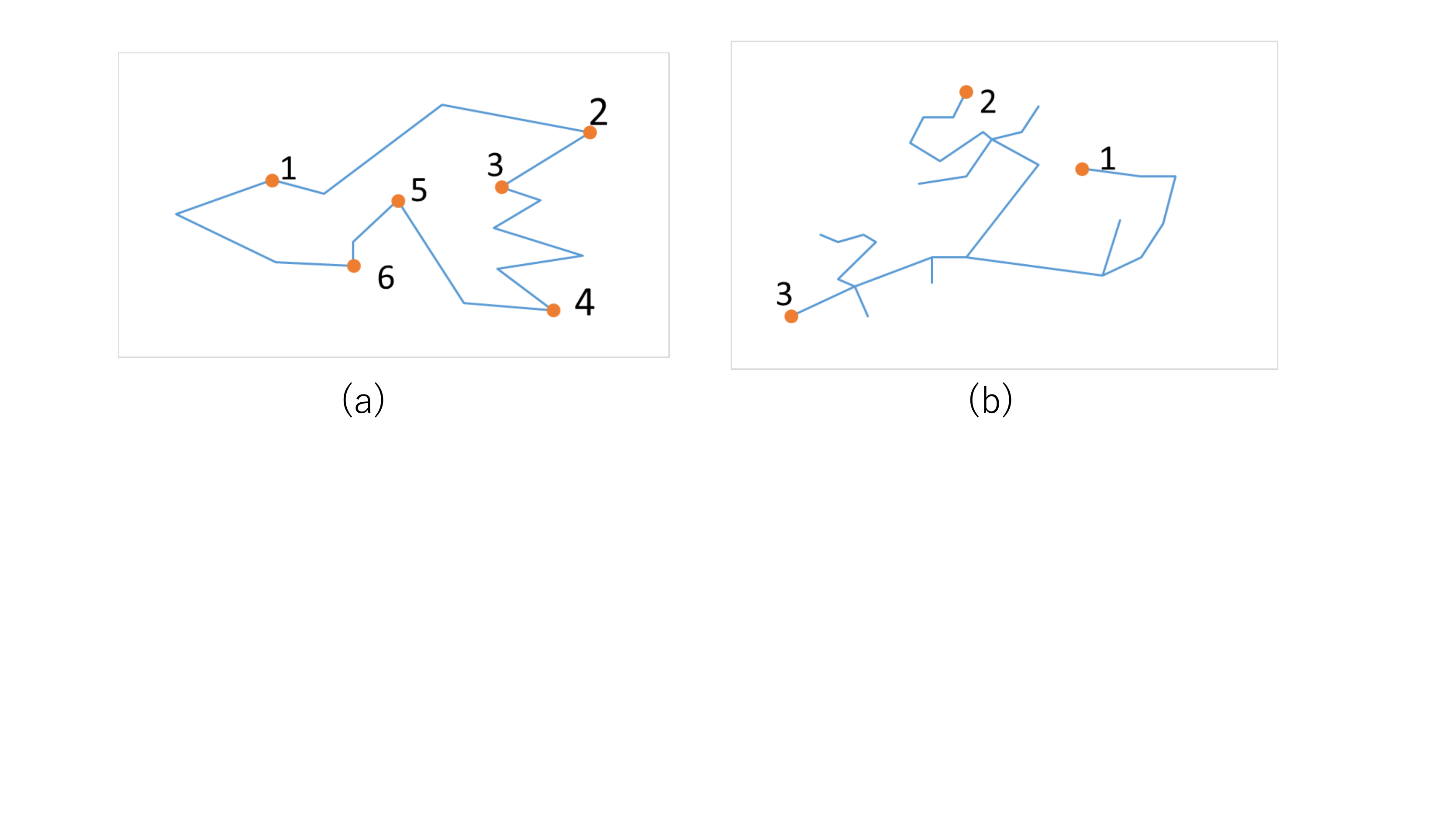}
\caption{Realistic network model.}
\label{realNet}
\end{center} 
\end{figure}

This subsection offers numerical examples of a realistic network model (Fig. \ref{realNet}), which was used in a previous study \cite{infocomSaito}.

The network in Fig. \ref{realNet} was used to evaluate Eq. (\ref{th_m1eq}) under the weakest arrangement. When $s,t$ are nodes 2 and 4, $\net$ becomes almost convex. Thus, Eq. (\ref{th_m1eq}) should be satisfied. In addition, when $s,t$ are nodes 2 and 3, Eq. (\ref{th_m1eq}) should also be satisfied because of Corollary \ref{col0}. For other cases, Eq. (\ref{th_m1eq}) becomes an approximation.

\begin{figure}[tbh]
\begin{center} 
\includegraphics[width=8cm]{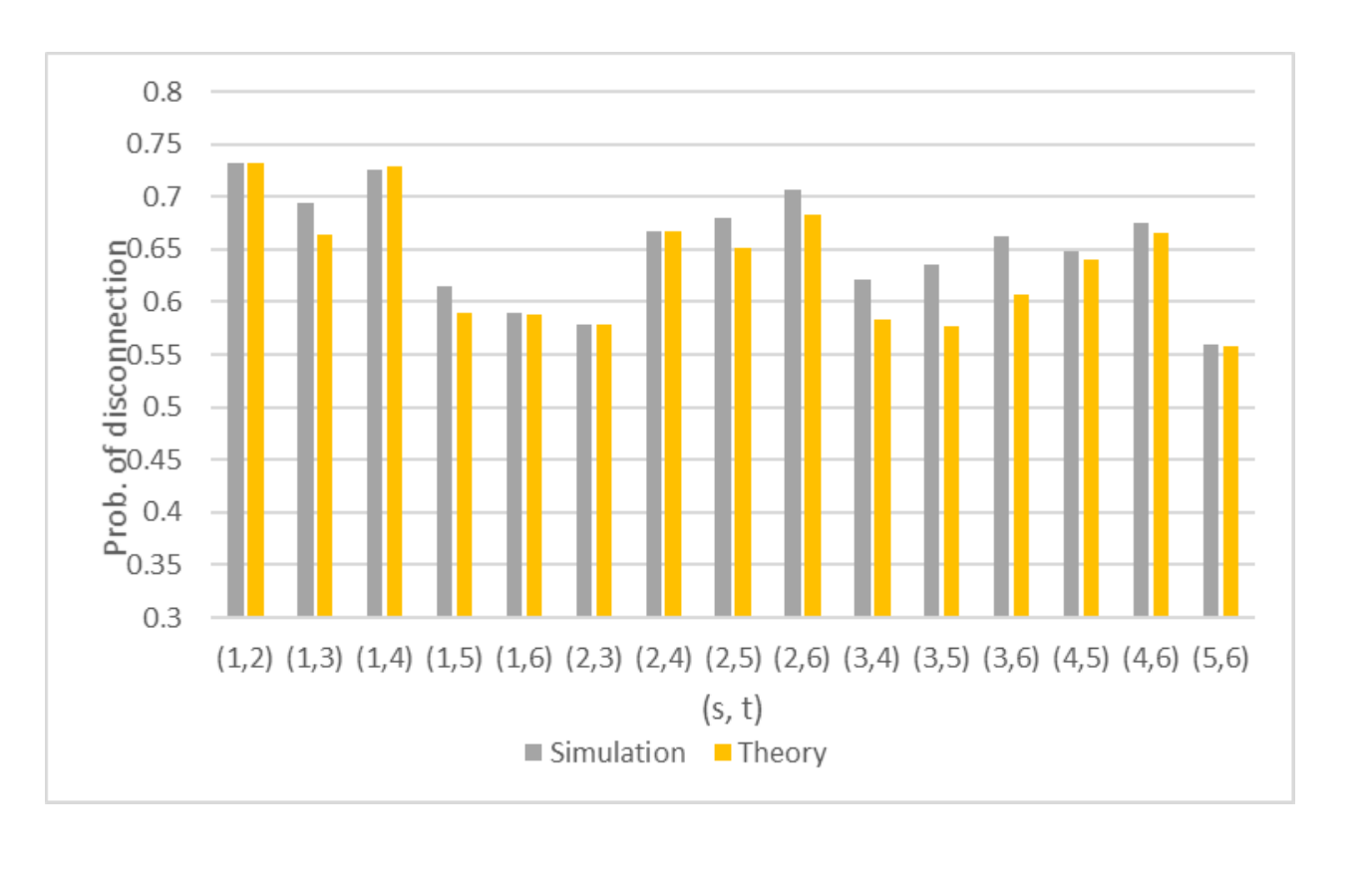}
\caption{$\Pr(s\discon t)$ for network in Fig. \ref{realNet}.}
\label{real_result1}
\end{center} 
\end{figure}

The evaluation results of Eq. (\ref{th_m1eq}) are plotted in Fig. \ref{real_result1}. As expected, simulation and theory showed good agreement when $(s, t)$ were (2, 3) and (2,4).  When $(s, t)$ were (3, 5) and (3, 6), their differences became large. This seems to be the result of disconnecting $\outer_1$ and $\outer_2$ due to disconnection between nodes 2 and 4 often occurring when $(s,t)=(3,5),(3,6)$, but Eq. (\ref{th_m1eq}) does not take into account such a disconnection. The reason for the poor approximation accuracy when $(s, t)=$(1, 3), (2, 5), (2, 6), and (3, 4) seems to be the same.
However, the simulation and theoretical results showed good agreement as a whole, even when the assumptions were not satisfied.

\section{Conclusion}
This paper introduced the disaster-endurance (DE) levels of a network that are specified by, for example, protection mechanisms and materials of ducts that contain optical fiber cables. The probability of connecting two given nodes was analyzed under multiple DE levels and the following results were obtained.

1. The probability of connecting two given nodes was derived in a closed form under the weakest arrangement.  It is the maximum lower bound of that probability under any arrangement.

2. When a part of a network can be protected, the placement of that part was determined to decrease that probability.  It was shown that protection around end nodes and inner parts of routes was critical.

Although this paper assumed $D=R_G$, analyses assuming a more generic disaster remain as future study.

\section*{Acknowledgment}
This work was supported by KAKENHI JSPS Grant Number JP-21K11864.



\begin{thebibliography}{99}
\bibitem{AsBeSo11} E. Asimakopoulou, N. Bessis, S. Sotiriadis, F. Xhafa, and L. Barolli, A Collective Intelligence Resource Management Dynamic Approach for Disaster Management: A Density Survey of Disasters Occurrence, IEEE INCoS, pp. 735-740, 2011.
\bibitem{ntt_commag} Mitsuyoshi Kobayashi, Experience of Infrastructure Damage Caused by the Great East Japan Earthquake and Countermeasures against Future Disasters, IEEE Communications Magazine, pp. 23-29, March 2014.
\bibitem{china} Y. Ran, Considerations and Suggestions on Improvement of Communication Network Disaster Countermeasures after the Wenchuan Earthquake, IEEE Communications Magazine, pp. 44-47, January 2011.
\bibitem{survey0}M. F. Habib et al., Disaster survivability in optical communication networks, Computer Communications, 36, pp. 630-644, 2013.
\bibitem{ntt} https://www.ntt.co.jp/saitai/en/3principles.html (Accessed on January 8, 2020).
\bibitem{commag_article} B. Mukherjee, M. F. Habib, and F. Dikbiyik, Network adaptability from disaster disruptions and cascading failures, IEEE Communications Magazine, pp. 230-238, May 2014.
\bibitem{disaster-free}Hiroshi Saito, Concept and Implementation of Disaster-free Network, Keynote speech, DRCN2015, Kansas City, 2015.
\bibitem{INFOCOMdisaster_avoidance}H. Saito, H. Honda and R. Kawahara, Disaster Avoidance Control against Heavy Rainfall, INFOCOM, 2017.
\bibitem{ToN_disaster_avoidance} H. Honda and H. Saito, Nation-Wide Disaster Avoidance Control Against Heavy Rain, IEEE Trans. Networking, 27, 3, pp. 1084-1097, 2019.
\bibitem{ToN_saito} Hiroshi Saito, Analysis of Geometric Disaster Evaluation Model for Physical Networks, IEEE Trans. Networking, 23, 6, pp. 1777-1789, 2015.
\bibitem{JLT_saito} Hiroshi Saito, Spatial Design of Physical Network Robust against Earthquakes, IEEE/OSA Journal of Lightwave Technology, 33, 2, pp. 443-458, 2015.
\bibitem{Commag_Tran} P. N. Tran and H. Saito, Geographical Route Design of Physical Networks Using Earthquake Risk Information, IEEE Communication Magazine, 54, 7, pp. 131-137, 2016.
\bibitem{JLT_Tran} P. N. Tran and H. Saito, Enhancing Physical Network Robustness against Earthquake Disasters with Additional Links, IEEE/OSA Journal of Lightwave Technology, 34, 22, pp. 5226-5238, 2016. 
\bibitem{infocomSaito} Hiroshi Saito, Geometric evaluation of survivability of disaster-affected network with probabilistic failure, in Proc. IEEE Conf. Comput. Commun. (INFOCOM), pp. 1608-1616, 2014.
\bibitem{zukerman_taiwan} Mingbo Zhao, Tommy W. S. Chow, Peng Tang, Zengfu Wang, Jun Guo, and Moshe Zukerman, Route Selection for Cabling Considering Cost Minimization and Earthquake Survivability Via a Semi-Supervised Probabilistic Model, IEEE Transactions on Industrial Informatics, 13, 2, pp. 502-511, 2017.
\bibitem{protection}Zengfu Wang, Qing Wang, Moshe Zukerman, Bill Moran, A Seismic Resistant Design Algorithm for Laying and Shielding of Optical Fiber Cables, IEEE/OSA Journal of Lightwave Technology, 35, 14, pp. 3060-3074, 2017. 
\bibitem{second_type}Hiroshi Saito, Theoretical Design of Geographical Route of Communications Network Supplied by Power Grid to Minimize Disaster Damage, submitted for publication.
\bibitem{underseaCableFailure}W. Wu, B. Moran, J. Manton, and M. Zukerman, Topology design of undersea cables considering survivability under major disasters, WAINA, pp. 1154-1159, 2009.
\bibitem{light} C. Cao, M. Zukerman, W. Wu, J. H. Manton, and B. Moran, Survivable topology design of submarine networks, IEEE J. Lightwave Technology, 31, 5, pp. 715-730, 2013.
\bibitem{cqr} M. T. Gardner and C. Beard, Evaluating geographic vulnerabilities in networks, Communications Quality and Reliability (CQR), pp. 1-6, 2011.
\bibitem{failureToN} S. Neumayer, G. Zussman, R. Cohen, and E. Modiano, Assessing the vulnerability of the fiber infrastructure to disasters, IEEE/ACM Trans. Networking, 19, 6. pp. 1610-1623, 2011.
\bibitem{failureINFOCOM} S. Neumayer and E. Modiano, Network reliability with geographically correlated failures, IEEE INFOCOM, 2010, pp. 1-9.
\bibitem{OR} D. Bienstock, Some generalized max-flow min-cut problems in the plane, Math. Oper. Res., 16, 2, pp. 310-333, 1991.
\bibitem{sen} A. Sen, B. H. Shen, L. Zhou, and B. Hao, Fault-tolerance in sensor networks: A new evaluation metric, IEEE INFOCOM, 2006, pp. 1-12.
\bibitem{discMinCut} S. Neumayer, A. Efrat, and E. Modiano, Geographic max-flow and min-cut under a circular disk failure model, IEEE INFOCOM, 2012, pp. 2736-2740.
\bibitem{wdmFailure} P. Agarwal, A. Efrat, S. Ganjugunte, D. Hay, S. Sankararaman, and G. Zussman, The resilience of WDM networks to probabilistic geographical failures, in Proc. IEEE INFOCOM, 2011, pp. 1521-1529.
\bibitem{icct} Y. Zhang, J. Y. Wang, and W. Li, Assessing the safety risk grade of optical network from the network geography distribution, Communication Technology (ICCT), pp. 709-712, 2012.
\bibitem{3pointFit}S. Trajanovski, F. A. Kuipers, A. Ili\'{c}, J. Crowcroft, and P. Van Mieghem, Finding critical regions and region-disjoint paths in a network, IEEE/ACM Trans. Networking, to be published.
\bibitem{Santalo} L. A. Santal\'o, Integral Geometry and Geometric Probability, Second edition. Cambridge University Press, Cambridge, 2004. 
\bibitem{saitoHP} http://www9.plala.or.jp/hslab/supplement/integral\_geo.pdf (Accessed on December 17, 2020).
\bibitem{earthquake} http://www.j-shis.bosai.go.jp/map/?lang=en (Accessed on December 24, 2019).
\end{thebibliography}
\end{document}